\documentclass[
  aps,
  prl,
  reprint,
  superscriptaddress,
  nofootinbib
]{revtex4-2}

\usepackage{amsmath,amssymb}
\usepackage{xcolor}
\usepackage{bm}
\usepackage{graphicx}
\usepackage{dcolumn}
\usepackage{float}
\usepackage{physics}
\usepackage{comment}
\usepackage{subcaption}
\usepackage[font=small,labelfont=bf]{caption}
\usepackage{hyperref}

\begin{document}

\title{Semiclassical scaling of eigenstate thermalization in single-particle chaotic systems}

\author{Yaoqi Ye}
\email{yyqnmcae@gmail.com}
\affiliation{Department of Modern Physics, University of Science and Technology of China, Hefei 230026, China}
\affiliation{Max Planck Institute for the Physics of Complex Systems, Nöthnitzer Str. 38, 01187 Dresden, Germany}

\author{Xiao Wang}
\email{wx2398@ustc.edu.cn}
\affiliation{Wilczek Quantum Center, Shanghai Institute for Advanced Studies, University of Science and Technology of China, Shanghai 201315, China}
\affiliation{Department of Modern Physics, University of Science and Technology of China, Hefei 230026, China}
\affiliation{
CAS Key Laboratory of Microscale Magnetic Resonance, University of Science and Technology of China, Hefei 230026, China}

% Add or remove authors and affiliations as needed.
% \author{Second Author}
% \affiliation{Department of Physics, University Name, City, Postal Code, Country}

\date{\today}

\begin{abstract}
We study the off-diagonal matrix elements of real-space observables in time-reversal-invariant single-particle chaotic systems. By analyzing the semiclassical expression for the off-diagonal variance derived from Berry's conjecture, we show that the banded structure of the observable matrix emerges naturally. For local observables, we identify a characteristic bandwidth associated with a late-time timescale inversely proportional to the particle velocity. We further show that, for systems with steep-wall confinement, the predicted magnitude follows the entropy scaling of the eigenstate thermalization hypothesis (ETH), multiplied by an additional kinetic-energy-dependent factor that is independent of spatial dimension and is not captured by conventional many-body ETH. We illustrate these results through a case study of quantum billiards and verify the semiclassical scaling numerically in a generalized quarter-Sinai billiard. Our results elucidate the dynamical implications of Berry's conjecture and provide a comparison between single-particle eigenstate thermalization and many-body ETH.
\end{abstract}

\maketitle

\textit{Introduction.}---
Understanding how an isolated quantum system approaches thermal equilibrium
from a nonequilibrium initial state is a central problem at the interface of
quantum mechanics and statistical physics. A widely used framework for
addressing this problem is the eigenstate thermalization hypothesis
(ETH)~\cite{Deutsch_1991,PhysRevE.50.888,Srednicki_1999}, which characterizes the statistical properties of matrix elements of observables in the eigenbasis of chaotic systems. For a physical
observable $\hat O$, ETH postulates that its matrix elements in the energy
eigenbasis take the form
\begin{equation}
  \langle E_i|\hat O|E_j\rangle
  =
  O_{\mathrm{mc}}(\bar E)\delta_{ij}
  +
  e^{-S(\bar E)/2}
  f_O(\bar E,\omega)r_{ij},
  \label{eq:eth_offdiag}
\end{equation}
where $\bar E=(E_i+E_j)/2$,
$\omega=E_i-E_j$, and $S(E)$ is the thermodynamic entropy at energy  $E$. The variables $r_{ij}$ have zero mean and unit variance within an
appropriate local spectral window. The smooth diagonal function
$O_{\mathrm{mc}}(\bar E)$ determines equilibrium expectation values, whereas
the off-diagonal envelope $f_O(\bar E,\omega)$ controls relaxation and dynamical correlation functions
\cite{Feingold_1986,D'Alessio03052016}. 

The general structure of the off-diagonal envelope has attracted considerable attention in recent years. Random matrix theory has shown that orthogonality-induced eigenstate correlations are essential for reproducing the ETH structure of off-diagonal matrix elements and their fluctuations \cite{nation2018off}. Beyond random-matrix considerations, the low-frequency structure and correlations of off-diagonal ETH matrix elements have been related to autocorrelation decay, shown to encode diffusive, subdiffusive, and superdiffusive transport, and used to constrain the crossover to random-matrix behavior \cite{luitz2016anomalous,PhysRevB.103.235137,dymarsky2022bound,PhysRevX.15.011059}. Complementarily, semiclassical approaches based on eigenfunction statistics and operator Weyl symbols have derived analytical structures for the diagonal and off-diagonal ETH functions and connected the off-diagonal bandwidth to classical phase-space dynamics \cite{xbtc-hlxw,xjc4-238b}.

In this Letter, we study the off-diagonal matrix elements of real-space
observables in time-reversal-invariant single-particle chaotic systems. While previous studies of ETH have mainly focused on many-body lattice models~\cite{PhysRevE.82.031130,rigol2008thermalization,PhysRevLett.103.100403,PhysRevA.80.053607,PhysRevE.87.012118,PhysRevLett.111.050403,PhysRevE.89.042112,PhysRevA.90.033606,PhysRevLett.112.130403,PhysRevE.90.052105,PhysRevE.91.012144,PhysRevE.93.032104,PhysRevB.93.134201,PhysRevE.96.012157,PhysRevLett.120.200604,PhysRevLett.122.070601,PhysRevB.99.155130,PhysRevE.100.062134,PhysRevLett.124.040603,PhysRevLett.125.070605,PhysRevB.102.075127,PhysRevLett.125.050603,PhysRevE.102.042127,PhysRevE.102.062113,PhysRevLett.126.120602,PhysRevE.103.012129,PhysRevB.103.235137,PhysRevE.103.062133}, eigenstate thermalization in the single-particle sector of quantum-chaotic quadratic Hamiltonians has been addressed in Ref.~\cite{PhysRevB.104.214203}. A recent study of the Feingold--Peres model~\cite{94k1-8w3n} has also investigated the connection between few-body quantum chaos and ETH. Here, instead of analyzing lattice models, we study a general class of continuum single-particle systems with time-reversal symmetry. In the paradigmatic setting of chaotic billiards, the statistics of diagonal observable matrix elements and the scaling of their variance have been investigated in Ref.~\cite{Barnett_2006}. Here, we study the off-diagonal matrix elements of real-space observables. Based on Berry's conjecture, we derive an analytical expression for the variance of off-diagonal matrix elements, from which the banded structure of the observable matrix emerges naturally. For local observables, we find that the bandwidth is proportional to $\hbar$ and the velocity, consistent with the inverse timescale of the corresponding classical dynamics. For systems with steep-wall confinement, the magnitude exhibits an explicit power-law scaling with kinetic energy. This scaling can be decomposed into an entropy-dependent factor, analogous to that appearing in the many-body ETH, and a kinetic-energy-dependent factor that is independent of spatial dimension. 
These results elucidate the general dynamical implications of Berry's conjecture in the single-particle case.

Following our general semiclassical analysis, we examine chaotic quantum billiards, which provide a particularly clean setting for testing our predictions for several reasons:
First, classical ergodicity has been rigorously established for dispersing billiards \cite{Sinai_1970,Bunimovich_1979}. Second, quantum billiards are paradigmatic quantum-chaotic systems whose level statistics agree with the predictions of random matrix theory ~\cite{PhysRevLett.42.1189,PhysRevLett.52.1,PhysRevA.37.3067}. For a generalized quarter-Sinai billiard, we numerically verify our semiclassical scaling expression and further find that the normalized off-diagonal elements follow a standard Gaussian distribution within an appropriately defined scaling window at high energies, consistent with the predictions of many-body ETH.

\textit{Notation.}---
We introduce the notation used throughout this work. We use polar coordinates
in the integrations below. Vectors are denoted by bold symbols, such as
$\bm{q}$, whereas their magnitudes are denoted by the corresponding nonbold
symbols, such as $q$. The angular coordinates on the unit sphere $S^{d-1}\subset\mathbb{R}^d$ are denoted by $\bm{\theta}_{d-1}$, with solid-angle element $d\bm{\theta}_{d-1}$ and total solid angle $S_{d-1}=\int_{S^{d-1}}d\bm{\theta}_{d-1}$.

The quantum momentum and position operators are denoted by
$\hat{\bm{p}}$ and $\hat{\bm{q}}$, with respective eigenstates
$\lvert \bm{p}\rangle$ and $\lvert \bm{q}\rangle$. Their classical
counterparts are denoted by $\bm{p}$ and $\bm{q}$. The Hamiltonian is written
as $H(\hat{\bm{p}},\hat{\bm{q}})$, with energy eigenstates
$\lvert E_i\rangle$ ordered by energy, and the observable of interest is
denoted by $O(\hat{\bm{p}},\hat{\bm{q}})$. The corresponding classical
functions, $H(\bm{p},\bm{q})$ and $O(\bm{p},\bm{q})$, are obtained by
replacing the quantum operators $(\hat{\bm{p}},\hat{\bm{q}})$ with the
classical variables $(\bm{p},\bm{q})$.

The Wigner function associated with the energy eigenstate
$\lvert E_i\rangle$ is defined as
\begin{equation}
W_i(\bm{p},\bm{q})
:=
\frac{1}{(2\pi\hbar)^d}
\int d\bm{r}\,
\psi_i^{*}\left(\bm{q}+\frac{\bm{r}}{2}\right)
\psi_i\left(\bm{q}-\frac{\bm{r}}{2}\right)
e^{i\bm{p}\cdot\bm{r}/\hbar},
\end{equation}
where $d$ is the dimension of the configuration space and
$\psi_i(\bm{q})=\langle\bm{q}\vert E_i\rangle$ is the corresponding energy
eigenfunction in the position representation.

\textit{Semiclassical expression for single-particle systems.}---
We consider a general $d$-dimensional single-particle system with chaotic
classical dynamics, where $d\geqslant 2$. The quantum Hamiltonian is written
as
\begin{equation}
  \hat{H}=\frac{\hat{\bm{p}}^{2}}{2m}+V(\hat{\bm q}),
  \label{eq:general_H}
\end{equation}
where $V(\hat{\bm q})$ is a potential for which the corresponding classical
dynamics is chaotic.

For this class of systems, we consider the semiclassical regime defined by~\cite{Berry_Mount_1972}
\begin{equation}
  \xi_{sm}^{-1}\equiv \sqrt{2m(E-V(\bm{q}))}/\hbar\gg \xi_V^{-1},
  \label{eq:semiclassical_limit}
\end{equation}
throughout most of the classically allowed region, where $\xi_V$ is the
spatial scale over which the potential varies.

Berry's conjecture is expected to hold in this limit. According to the
conjecture, the ensemble-averaged Wigner function of the $i$th eigenstate
$\lvert E_i\rangle$ is approximately uniform over the corresponding
classical energy shell:
\begin{equation}
  \overline{W_i}(\bm p,\bm q)
  \simeq
  \delta(E_i-H(\bm p,\bm q))/\mathcal{N}(E_i),
  \label{eq:wigner_function}
\end{equation}
where
$\mathcal{N}(E_i)=\int d\bm{q}\,d\bm{p}\,
\delta(E_i-H(\bm p,\bm q))$.
The overline denotes an average over a fictitious ensemble of eigenstates.
Equivalently, it may be interpreted as an average over narrow energy
windows centered at $E_i$ and $E_j$, respectively, because the eigenfunctions are expected to become highly irregular in
the semiclassical limit~\cite{PhysRevLett.97.214101,Berry_1977}.

Using Eq.~\eqref{eq:wigner_function}, one can derive an analytical
expression for the variance of the off-diagonal matrix elements of an
arbitrary observable
$O(\hat{\bm{p}},\hat{\bm{q}})$~\cite{xjc4-238b}. For observables that depend only on
the position operator,
$O(\hat{\bm{p}},\hat{\bm{q}})=O(\hat{\bm{q}})$, this expression can be
simplified exactly as follows ~\cite{supplemental}:
\begin{widetext}
\begin{equation}
  \begin{aligned}
  &\overline{|\bra{E_i}O(\hat{\bm{q}})\ket{E_j}|^2}
  =
  \hbar^{d-2}
  \bigg(\frac{2}{m}\bigg)^{d/2-1}
  \Gamma^2\bigg(\frac{d}{2}\bigg){A(E_i)}^{-1}{A(E_j)}^{-1}
  \int d\bm{q}\,d\tilde{q}\,
  J_{d/2-1}\bigg(
  \frac{\sqrt{2m(E_i-V(\bm{q}))}\tilde{q}}{\hbar}
  \bigg)
  \\
  &\times J_{d/2-1}\bigg(
  \frac{\sqrt{2m(E_j-V(\bm{q}))}\tilde{q}}{\hbar}
  \bigg)
  \big[(E_i-V(\bm{q}))(E_j-V(\bm{q}))\big]^{
  \left(\frac{d}{4}-\frac{1}{2}\right)}
  \int d\bm{\theta}_{d-1}\,
  O(\bm{q}+\tilde{\bm{q}}/2)
  O(\bm{q}-\tilde{\bm{q}}/2)
  \tilde{q}.
  \end{aligned}
  \label{eq:offdiagnal variance}
\end{equation}
\end{widetext}
where $A(E)=\int_{V(\bm{q})<E} d\bm{q}\,[E-V(\bm{q})]^{d/2-1}$, $J_\nu$ denotes the Bessel function arising from the two-point
correlation function of the wavefunction, and $\Gamma$ is the Gamma function. The integration domain for
$\bm{q}$ is restricted to
$V(\bm{q})<\min\{E_i,E_j\}$.

Eq.~\eqref{eq:offdiagnal variance} predicts a banded structure for the
observable matrix. As the energy difference $|E_i-E_j|$ increases, the two
Bessel functions oscillate with increasingly different frequencies. Their
product therefore dephases rapidly, reducing its contribution to the
integral and suppressing the off-diagonal variance.

In the semiclassical limit defined by
Eq.~\eqref{eq:semiclassical_limit}, the Bessel function can be approximated
by
\begin{equation}
  J_{d/2-1}(x)
  \approx
  \sqrt{\frac{2}{\pi x}}
  \cos\bigg(x-\frac{\pi(d-1)}{4}\bigg).
  \label{eq:J_cos}
\end{equation}
Substituting this approximation into
Eq.~\eqref{eq:offdiagnal variance} yields
\begin{equation}
  \begin{aligned}
  &\overline{|\bra{E_i}O(\hat{\bm{q}})\ket{E_j}|^2}
  \approx
  \gamma{A(E_i)}^{-1}{A(E_j)}^{-1}\int d\bm{q}\,d\tilde{q}\,
  \\
  &\quad\times
  \cos\bigg(
  \frac{\sqrt{2m}\tilde{q}(E_i-E_j)}
  {\hbar\big(\sqrt{E_i-V(\bm{q})}
  +\sqrt{E_j-V(\bm{q})}\big)}
  \bigg)
  \\
  &\quad\times
  \big[(E_i-V(\bm{q}))(E_j-V(\bm{q}))\big]^{
  \frac{d}{4}-\frac{3}{4}}
  \\
  &\quad\times
  \int d\bm{\theta}_{d-1}\,
  O(\bm{q}+\tilde{\bm{q}}/2)
  O(\bm{q}-\tilde{\bm{q}}/2),
  \end{aligned}
  \label{eq:offdiagonal variance approx}
\end{equation}
where we define
\begin{equation}
  \gamma\equiv \frac{\hbar^{d-1}}{2\pi}\big(\frac{2}{m}\big)^{d/2-1/2}\Gamma^2\big(\frac{d}{2}\big)
  \label{eq:gamma}
\end{equation}
to simplify the expression.

We now consider a localized observable $O(\hat{\bm{q}})$ supported within a
region $\Omega_O$ away from the boundary of the classically allowed region.
Let its characteristic length scale $\xi_O$ satisfy
\begin{equation}
\xi_{sm}\ll\xi_O\leqslant \xi_V.
\label{eq:localized_condition}
\end{equation} 
Within $\Omega_O$, the kinetic energies can be
approximated as
\begin{equation}
E_i-V(\bm{q})
\approx
E_j-V(\bm{q})
\approx E_k,
\end{equation}
where $E_k$ is the mean kinetic energy in this region. It then follows from
Eq.~\eqref{eq:offdiagonal variance approx} that the characteristic bandwidth
$\omega_b$ of the observable matrix scales as
\begin{equation}
\omega_b
\propto
\hbar E_k^{1/2}m^{-1/2}
\propto
\hbar v,
\end{equation}
where $v=\sqrt{2E_k/m}$ is the characteristic particle velocity within the observable's support.

We next consider the magnitude of the off-diagonal variance. Conventional
many-body ETH predicts a scaling proportional to $e^{-S(E)}$. For the single-particle
system considered here, the classical microcanonical entropy is given by
\begin{equation}
  e^{S(E)}
  =
  \frac{S_{d-1}m^{d/2}}
  {2^{d/2+1}\pi^d\hbar^d}
  A(E).
  \label{eq:entropy}
\end{equation}
If the right-hand side of Eq.~\eqref{eq:entropy} is proportional to
$E_k^{d/2-1}$, which is the case for systems with steep-wall confinement, then
\begin{equation}
f_O(E,\omega)\simeq \hbar^{-1/2}E_k^{-1/4}g_O(\frac{\omega}{\hbar\sqrt{E_k}}).
\label{eq:f_O}
\end{equation}
This condition is satisfied
when $E-V(\bm{q})$ is approximately constant throughout most of the
classically allowed region and decreases rapidly to zero only near its
boundary. Such behavior occurs in systems with steep-wall confinement, such
as billiards. The $\hbar$ dependence of $|f_O(E,\omega)|^2$ ensures that its
Fourier transform into the time domain is independent of $\hbar$. The same dependence on the effective Planck constant was also obtained from a semiclassical analysis of the autocorrelation function in Ref.~\cite{94k1-8w3n}.

\textit{A simple model: chaotic billiards.}---
We now apply the semiclassical expression derived above to a
two-dimensional chaotic billiard with domain $\Omega$.

Consider a real-space observable of the scale-invariant form
\begin{equation}
  O(\hat{\bm{q}})
  =
  f(\hat{\bm{q}}/\sqrt{S_\Omega}),
\end{equation}
where $S_\Omega$ is the area of the billiard domain.

Using Eq.~\eqref{eq:offdiagonal variance approx}, we obtain the following
expression for the variance of the off-diagonal matrix elements:
\begin{equation}
  \begin{split}
  \overline{|\bra{E_i}O(\bm{\hat{q}})\ket{E_j}|^2}
  &\approx
  \frac{\hbar}
  {\pi\sqrt{2m}S_\Omega^{1/2}(E_iE_j)^{1/4}}
  \int d\bm{\widetilde{z}}\,
  \frac{\Phi(\bm{\widetilde{z}})}
  {\widetilde{z}}
  \\
  &\quad\times
  \cos\bigg[
  \frac{\sqrt{2mS_\Omega}}
  {\hbar(\sqrt{E_i}+\sqrt{E_j})}
  \widetilde{z}\omega
  \bigg],
  \end{split}
  \label{eq:Oq_offdiag_Phi_approx}
\end{equation}
where
$\omega=E_i-E_j$,
$\bm{z}=\bm{q}/\sqrt{S_\Omega}$, and
$\widetilde{\bm{z}}=\widetilde{\bm{q}}/\sqrt{S_\Omega}$. The function
$\Phi$ is defined by
\begin{equation}
  \begin{split}
  \Phi(\widetilde{\bm{z}})
  &=
  \int_{\Omega_0
  \cap(\Omega_0-\widetilde{\bm{z}}/2)
  \cap(\Omega_0+\widetilde{\bm{z}}/2)}
  d\bm{z}\,
  \\
  &\quad\times
  f\!\left(\bm{z}-\frac{\widetilde{\bm{z}}}{2}\right)
  f\!\left(\bm{z}+\frac{\widetilde{\bm{z}}}{2}\right).
  \end{split}
  \label{eq:Phi}
\end{equation}
Here $\Omega_0=\{\bm q/\sqrt{S_\Omega}:\bm q\in\Omega\}$ denotes the
rescaled billiard domain, which has unit area.

We now focus on the low-frequency window relevant to the pre-equilibrium
dynamics,
$|\omega|<\omega_m$, with
$\omega_m/\bar{E}\ll1$ and
$\bar{E}=(E_i+E_j)/2$. In this regime,
Eq.~\eqref{eq:Oq_offdiag_Phi_approx} reduces to
\begin{equation}
  \begin{split}
  &\overline{|\bra{E_i}O(\bm{\hat{q}})\ket{E_j}|^2}
  \approx
  \frac{\hbar}
  {\pi\sqrt{2m\bar{E}}S_\Omega^{1/2}}
  \bigg[
  1+\mathcal{O}\bigg(\frac{\omega^2}{\bar{E}^2}\bigg)
  \bigg]
  \\
  &\quad\times
  \int d\bm{\widetilde{z}}\,
  \frac{\Phi(\bm{\widetilde{z}})}{\widetilde{z}}
  \cos\bigg[
  \frac{\sqrt{mS_\Omega}}
  {\hbar\sqrt{2\bar{E}}}
  \widetilde{z}\omega\bigg(1
  +
  \mathcal{O}\bigg(\frac{\omega^2}{\bar{E}^2}\bigg)\bigg)
  \bigg].
  \end{split}
  \label{eq:Oq_offdiag_Phi_approx_w}
\end{equation}

This expression shows that the characteristic frequency scale governing the
variation of
$\overline{|\bra{E_i}O(\bm{\hat{q}})\ket{E_j}|^2}$ along the $\omega$ axis
is
\begin{equation}
  \omega_b\sim \frac{\hbar\sqrt{2\overline{E}}}{\sqrt{mS_\Omega}}
  =
  \frac{\hbar p}{m\sqrt{S_\Omega}},
  \label{eq:omega_b_billiard}
\end{equation}
for $|\omega|<\omega_m$, where
$p=\sqrt{2m\overline{E}}$.

We may therefore write the off-diagonal ETH envelope in the scaling form
\begin{equation}
  |f_O(E,\omega)|^2
  \sim
  \hbar^{-1}E^{-1/2}
  g_O^2\bigg(\frac{\omega}{\hbar\sqrt{E}}\bigg),
  \label{eq:f_O_sq}
\end{equation}
where $g_O$ is the scaling function.

In the semiclassical limit, the connected autocorrelation function of an
observable $O$ can be expressed as~\cite{D'Alessio03052016, PhysRevB.103.235137}
\begin{equation}
\begin{split}
  \langle O(t)O(0)\rangle_c
  &={}
  \int d\omega\,
  e^{\beta\omega/2-i\omega t/\hbar}
  \\
  &\times
  \left[
    |f_O(E,\omega)|^2
    +
    \frac{\omega}{2}
    \frac{\partial|f_O(E,\omega)|^2}{\partial E}
  \right],
\end{split}
  \label{eq:correlation_function_and_f}
\end{equation}
where
$\beta=\partial S(E)/\partial E=0$ for the two-dimensional billiards
considered here.

Using Eq.~\eqref{eq:Oq_offdiag_Phi_approx_w}, we obtain the symmetric part
of the connected autocorrelation function ~\cite{supplemental}:
\begin{equation}
  \langle O(t)O(0)\rangle_{\mathrm{sym}}
  =
  \frac{1}{2\pi}
  \int d\tilde{\bm{z}}\,
  \frac{\Phi\left(\tilde{\bm{z}}\right)}{\tilde{z}}
  \delta\left(
  \tilde{z}
  -
  \sqrt{\frac{2\bar{E}}{mS_\Omega}}|t|
  \right).
  \label{eq:symmetric_connected_correlation}
\end{equation}

Consequently, for the late-time dynamics resolved by the low-frequency scaling
form, the characteristic timescale obeys
$\tau\sim mp^{-1}\sqrt{S_\Omega}$. Because $p/m$ is the classical speed,
this scaling agrees with the classical traversal time of the billiard.

\begin{figure}[htbp]
  \centering
  \includegraphics[width=0.55\linewidth]{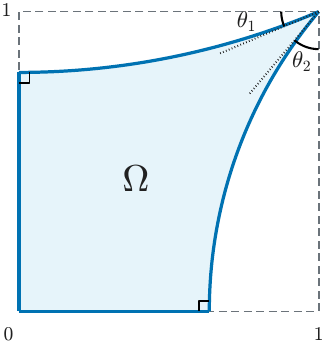}
  \caption{Geometry of the generalized quarter-Sinai billiard. The shaded domain $\Omega$ is bounded by two straight segments on the left and bottom and two circular arcs on the top and right. The angles $\theta_1=0.4$ and $\theta_2=0.7$ radians are measured between the arc tangents at their common vertex (dotted lines) and the horizontal and vertical directions, respectively. Each arc meets the adjacent straight boundary at a right angle.}
  \label{fig:generalized_sinai_geometry}
\end{figure}

\textit{Numerical results in a generalized quarter-Sinai billiard.}---
\label{sec:numerical_results}
We test the analytical predictions in a generalized quarter-Sinai billiard with concave walls (see Fig.~\ref{fig:generalized_sinai_geometry}). Classical ergodicity has been rigorously established for the full generalized Sinai billiard \cite{Sinai_1970}. And equivalently, we work in the odd-odd symmetry sector of the full billiard. We solve the eigenproblem for the generalized quarter-Sinai billiard using the scaling method proposed in Refs.~\cite{Barnett_2006,PhysRevE.52.2204}.

We set $\hbar=1$ and $m=1/2$, and calculate the matrix elements of the real-space observable $O(\bm q)=q_x$, where $q_x$ is the $x$-component of the position observable $\bm q$. This observable satisfies the localization condition in Eq.~\eqref{eq:localized_condition}, since $\xi_O=\xi_V\to\infty$ within the billiard domain.

\begin{figure*}[t]
  \centering
  \begin{subfigure}[t]{0.24\textwidth}
    \centering
    \includegraphics[width=\linewidth]{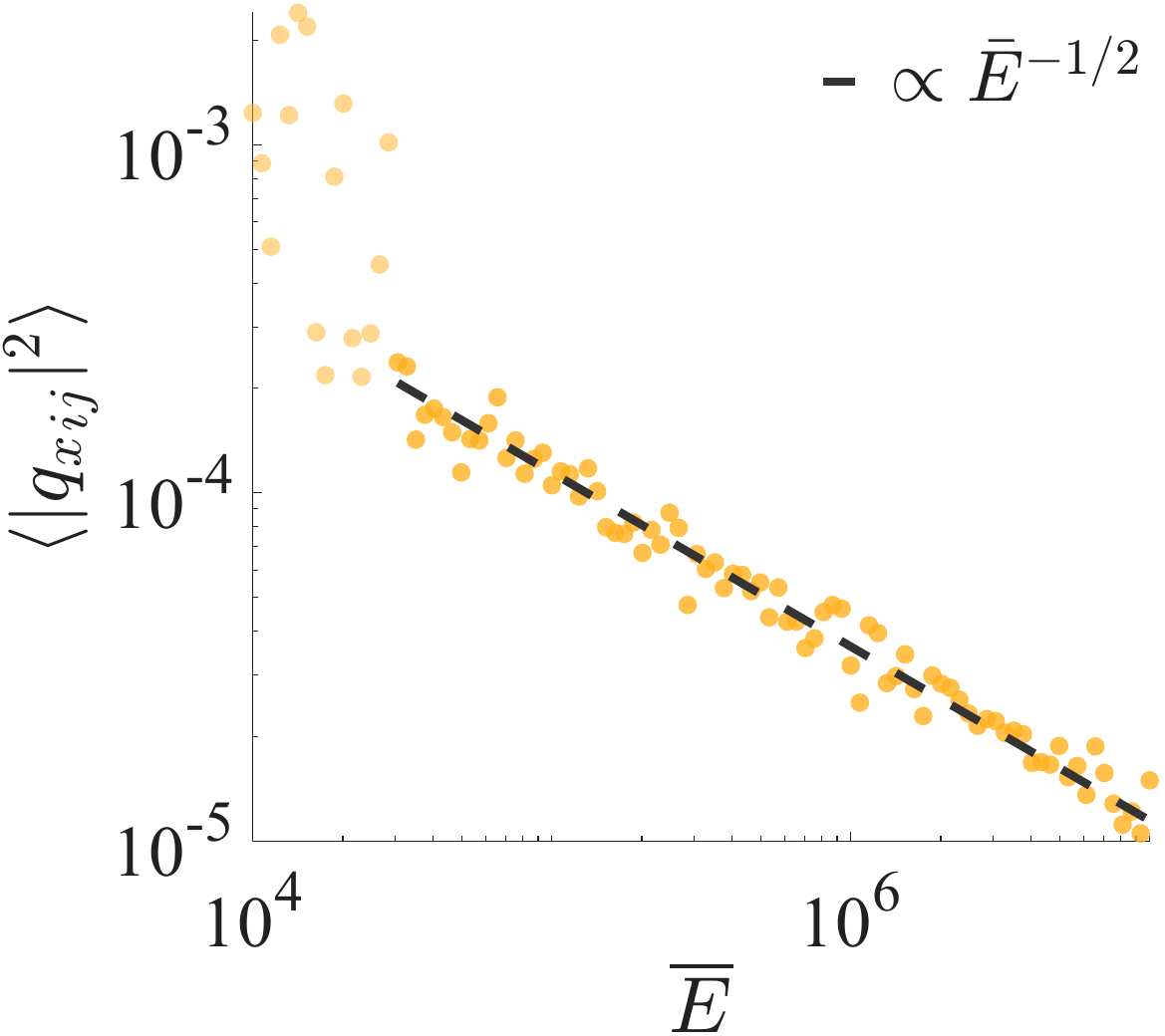}
    \caption{}
    \label{fig:generalized_sinai_targeted_u1}
  \end{subfigure}\hfill
  \begin{subfigure}[t]{0.24\textwidth}
    \centering
    \includegraphics[width=\linewidth]{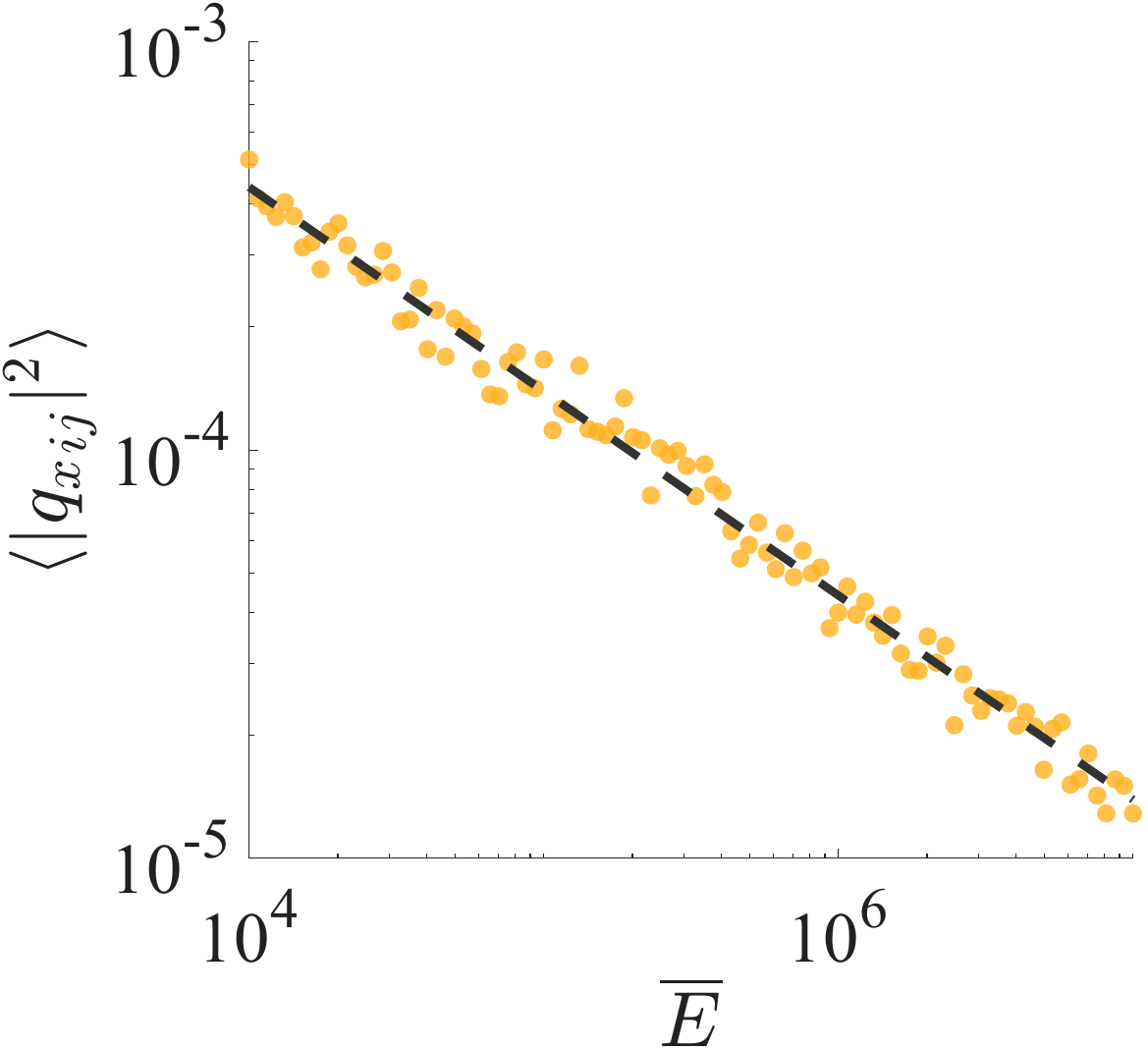}
    \caption{}
    \label{fig:generalized_sinai_targeted_u2}
  \end{subfigure}\hfill
  \begin{subfigure}[t]{0.24\textwidth}
    \centering
    \includegraphics[width=\linewidth]{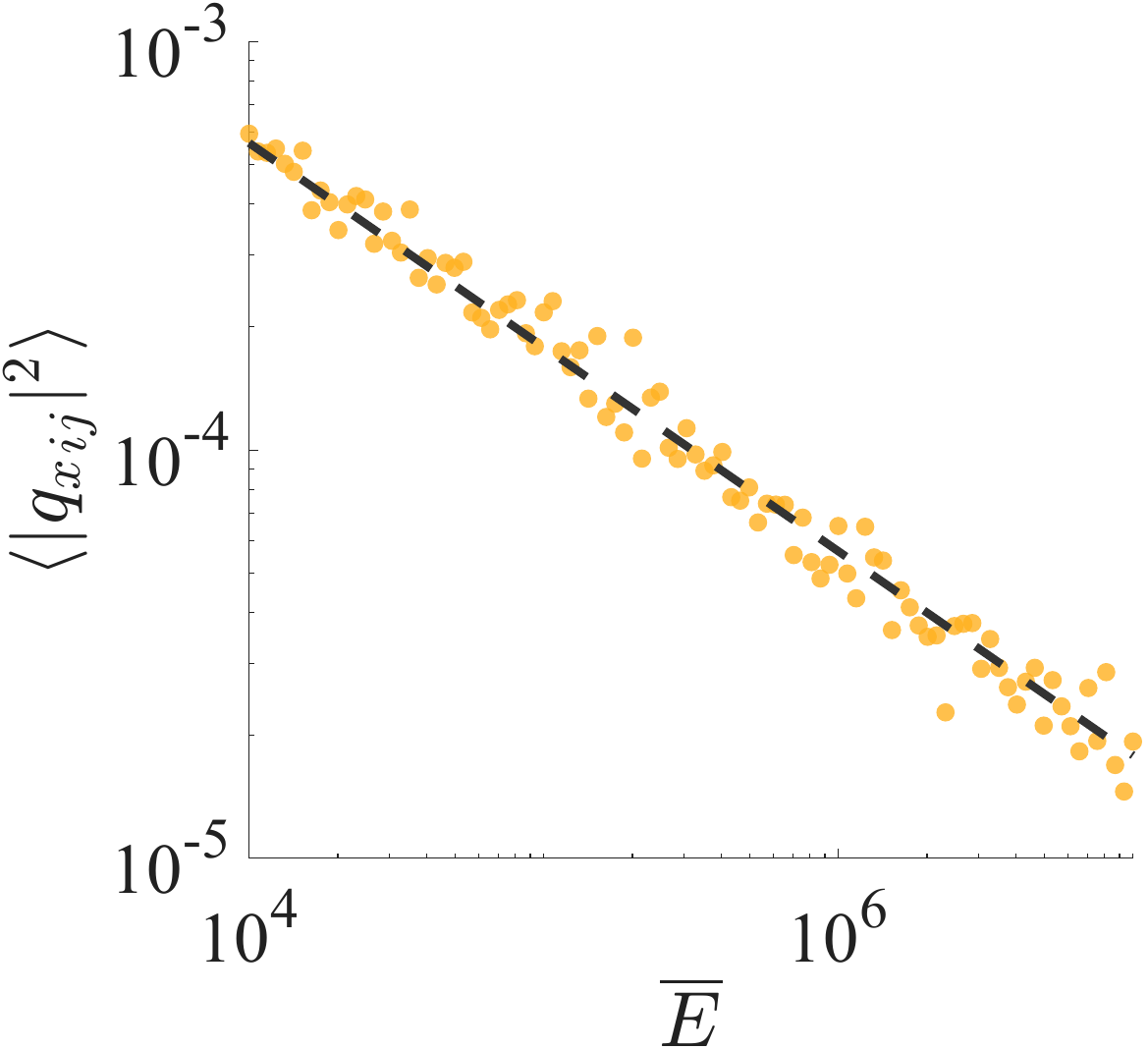}
    \caption{}
    \label{fig:generalized_sinai_targeted_u3}
  \end{subfigure}\hfill
  \begin{subfigure}[t]{0.24\textwidth}
    \centering
    \includegraphics[width=\linewidth]{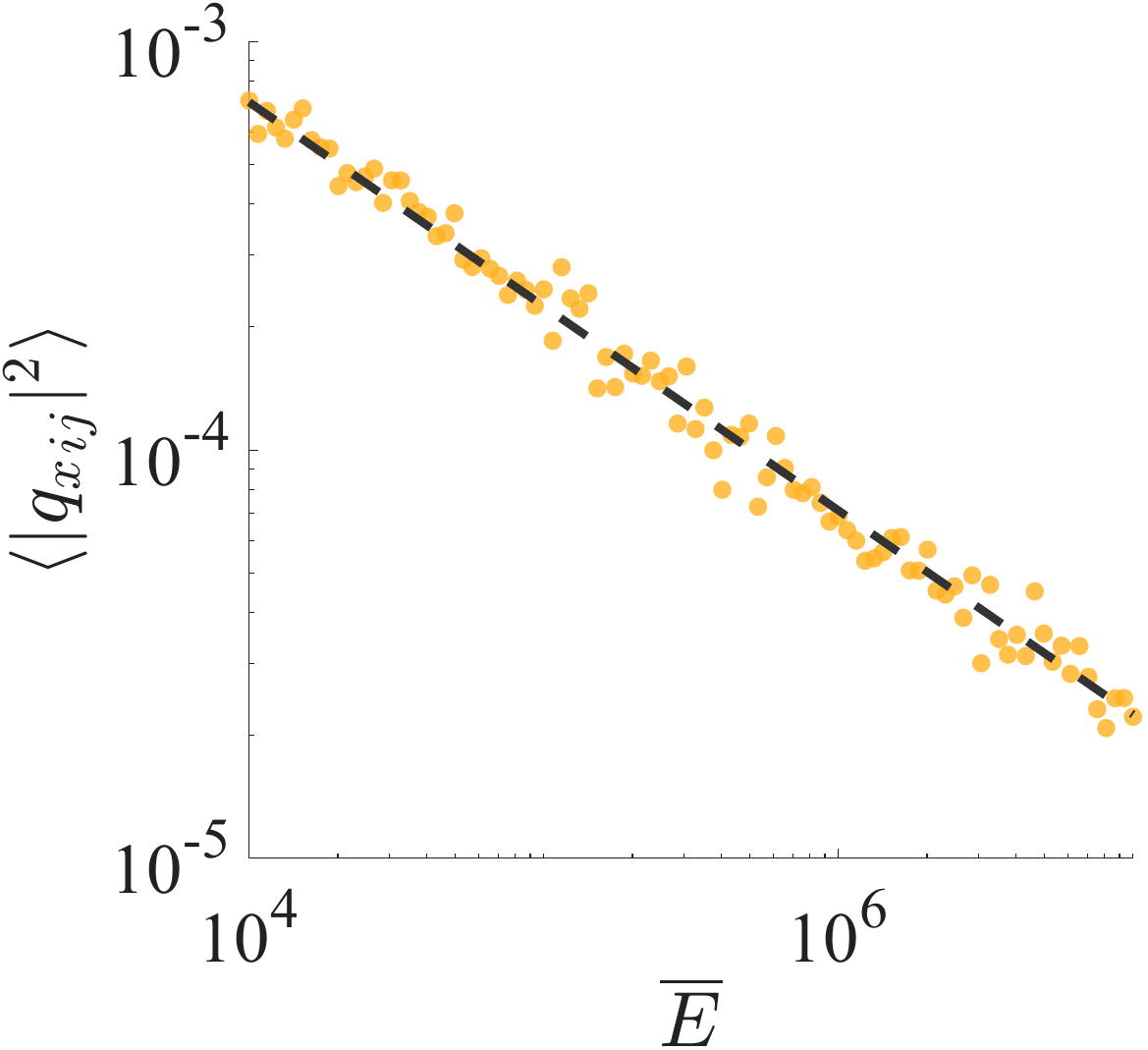}
    \caption{}
    \label{fig:generalized_sinai_targeted_u4}
  \end{subfigure}
  \caption{Semiclassical scaling of the off-diagonal variance along traces
  of fixed $u$. Panels (a)--(d) correspond to $u=1,2,3,4$, respectively.
  The vertical axis shows the local variance of the matrix elements
  ${q_x}_{ij}=\langle E_i|\hat q_x|E_j\rangle$.
  For each fixed $u$, we select 100 target mean energies $\bar E$
  logarithmically spaced between $10^4$ and $10^7$, with the corresponding
  energy separations set by $|\omega|=u\sqrt{\bar E}$.
  At each target point, the local variance is estimated by averaging
  $|{q_x}_{ij}|^2$ over the 100 eigenstate pairs closest to that point
  in energy space. The dashed black lines indicate the predicted semiclassical scaling, $\langle|{q_x}_{ij}|^2\rangle\propto\bar E^{-1/2}$, along each fixed-$u$ trace.}
  \label{fig:generalized_sinai_targeted_exponents}
\end{figure*}

We introduce the scaled energy separation
\begin{equation}
  u\equiv\frac{|\omega|}{\sqrt{\bar E}},
  \label{eq:alpha_definition}
\end{equation}
where $\bar E=(E_i+E_j)/2$ and $\omega=E_i-E_j$.
To verify the scaling form of the off-diagonal ETH envelope in Eq.~\eqref{eq:f_O_sq}, we test its power-law scaling with the mean energy $\bar E$ along traces of fixed $u$. Fig.~\ref{fig:generalized_sinai_targeted_exponents} shows good agreement with the predicted scaling for $u=2,3,4$ over the range $1\times 10^4<\bar E < 1\times 10^7$. For $u=1$, the semiclassical scaling emerges for $\bar E>3\times 10^4$ but is absent at lower energies due to finite-energy effects. The origin of the poorer agreement at smaller $u$ remains unclear.

\begin{figure}[htbp]
  \centering
  \includegraphics[width=\linewidth]{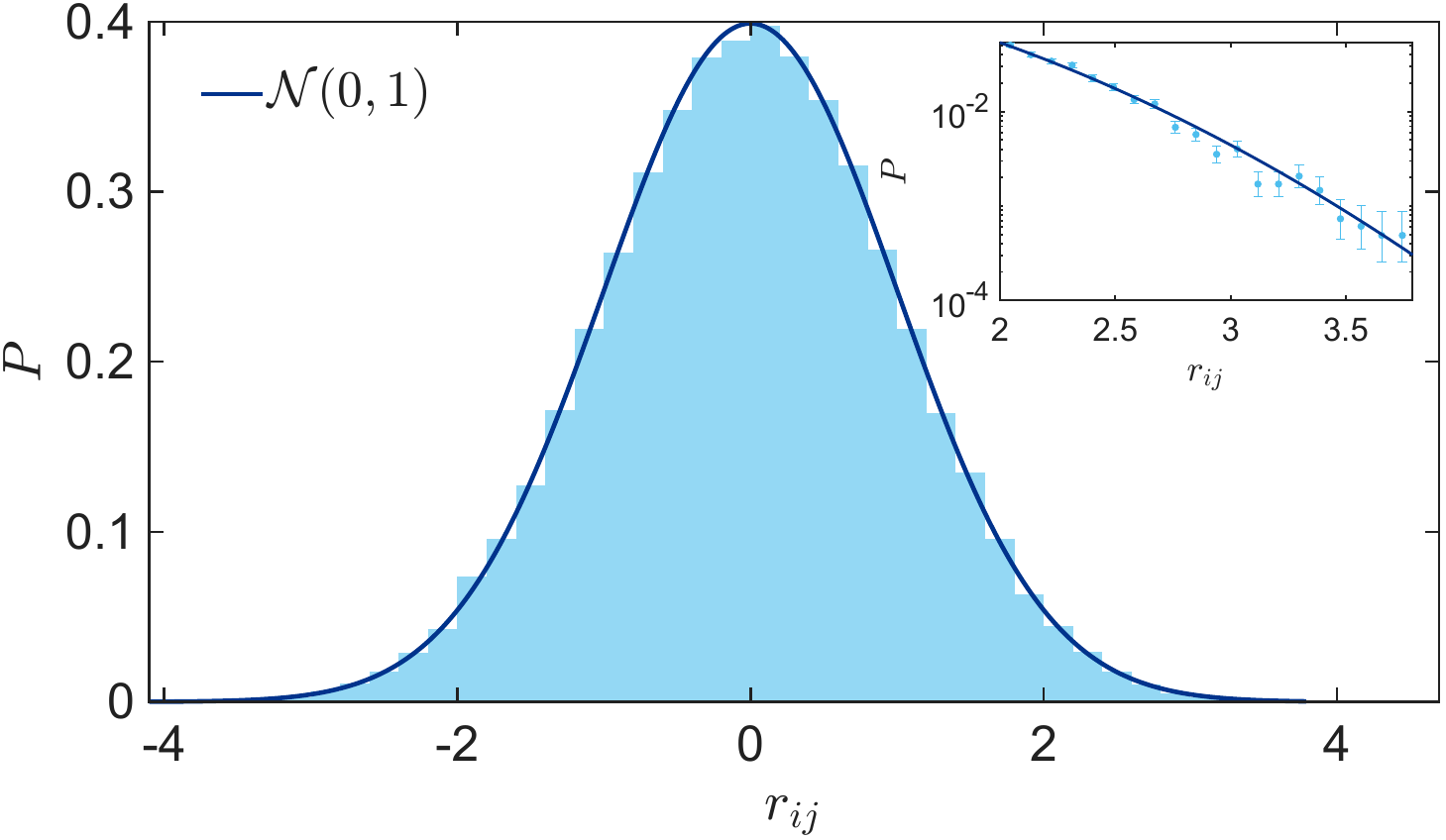}
  \caption{Probability density of the normalized off-diagonal matrix
  elements $r_{ij}$. The histogram contains $91\,695$ eigenstate pairs near
  $\bar E\simeq9\times10^6$ within the scaled energy-separation window
  $2\leq u=|\omega|/\sqrt{\bar E}\leq3$.
  The local variance is estimated by averaging $|{q_x}_{ij}|^2$ over a
  $9\times9$ block of eigenstate indices centered on $(i,j)$, excluding
  diagonal elements. The solid dark-blue curve shows the standard normal
  distribution $\mathcal{N}(0,1)$. The inset resolves the positive tail
  $r_{ij}\geq2$ on a logarithmic probability scale; points show the binned
  probability density, with error bars obtained from exact central
  $68.27\%$ Poisson confidence intervals for the bin counts.}
  \label{fig:generalized_sinai_scaled_eth_statistics}
\end{figure}

We also study the statistics of the off-diagonal matrix elements using eigenstates obtained by the scaling method. We define the normalized off-diagonal matrix elements as
\begin{equation}
r_{ij}
=
\frac{O_{ij}}
{\sqrt{\overline{|O_{ij}|^2}}},
\label{eq:normalized_eth_residual}
\end{equation}
where $\overline{|O_{ij}|^2}$ is the local variance and $O=q_x$ in our calculations. The distribution of $r_{ij}$ begins to approach a standard Gaussian at mean energies $\bar E\gtrsim 9\times10^6$ (see Fig.~\ref{fig:generalized_sinai_scaled_eth_statistics}). This behavior is consistent with Berry's random-wave conjecture, which models semiclassical eigenfunctions locally as superpositions of plane waves with uncorrelated phases~\cite{Berry_1977}, and with the predictions of conventional many-body ETH. At lower energies, the distribution of $r_{ij}$ deviates from Gaussian statistics, as an effect of finite energy.

\textit{Discussion and conclusion.}---
In this work, we analyzed the dynamical consequences of Berry's conjecture
by studying the variance of the off-diagonal matrix elements of real-space
observables in single-particle chaotic systems with time-reversal symmetry. By comparing the
semiclassical prediction with conventional many-body ETH, we showed that the
exponential entropy scaling provides a good approximation for systems with
steep-wall confinement in the low-frequency regime, apart from a dimension-independent kinetic-energy scaling that is not captured by
conventional ETH. Similar energy scaling may arise in other
classes of systems, a possibility
that we leave for future investigation.

Although the semiclassical prediction based on Berry's conjecture captures
the scaling properties of the off-diagonal variance well, a direct
comparison with the smoothed numerical variance reveals a systematic
quantitative discrepancy. This discrepancy persists even at $\bar E\simeq1.6\times10^7$, with no clear trend toward convergence to Eq.~\eqref{eq:offdiagnal variance} over the energy range studied in our numerical simulation. We suggest that this discrepancy may arise from correlations between
different eigenstates that are not captured by Berry's conjecture. 

\textit{Acknowledgments.}---
We thank Masudul Haque for helpful comments on the manuscript and Chengkai Lin, Jiaozi Wang, Wen-ge Wang, and Felix Fritzsch for useful discussions. Numerical simulations were performed on the computing cluster of the Max Planck Institute for the Physics of Complex Systems in Dresden.

\bibliography{references}

\end{document}

\title{\texorpdfstring{Supplemental Material for\\
``Semiclassical scaling of eigenstate thermalization in single-particle chaotic systems''}
{Supplemental Material for ``Semiclassical scaling of eigenstate thermalization in single-particle chaotic systems''}}

\author{Yaoqi Ye}
\email{yyqnmcae@gmail.com}
\affiliation{Department of Modern Physics, University of Science and Technology of China, Hefei 230026, China}
\affiliation{Max Planck Institute for the Physics of Complex Systems, N\"othnitzer Str. 38, 01187 Dresden, Germany}

\author{Xiao Wang}
\email{wx2398@ustc.edu.cn}
\affiliation{Wilczek Quantum Center, Shanghai Institute for Advanced Studies, University of Science and Technology of China, Shanghai 201315, China}
\affiliation{Department of Modern Physics, University of Science and Technology of China, Hefei 230026, China}
\affiliation{
CAS Key Laboratory of Microscale Magnetic Resonance, University of Science and Technology of China, Hefei 230026, China}

\date{\today}

\maketitle

% Use the prefix ``sm:'' for labels defined in this file, for example,
% \label{sm:eq:variance} and \label{sm:fig:phi}.

\section{Derivation of the off-diagonal variance in single-particle systems}
In this section, we derive
Eq.~\eqref{main:eq:offdiagnal variance} in the main text.

The variance of the off-diagonal matrix elements for a real-space observable can be written in the position representation as
\begin{equation}
  \overline{|\bra{E_i}O(\hat{\bm{q}})\ket{E_j}|^2}=\int d\bm{q}'d\bm{q}''\,\overline{\psi_i^{*}(\bm{q}'')\psi_i(\bm{q}')}\,\overline{\psi_j^{*}(\bm{q}')\psi_j(\bm{q}'')}O(\bm{q}')O(\bm{q}'').
  \label{SM:eq:offdiagonal variance init}
\end{equation}
We use the two-point correlation function derived from Berry's conjecture \cite{Berry_1977},
\begin{equation}
  \begin{aligned}
  \overline{\psi_i^{*}\big(\bm{q}-\frac{\tilde{\bm{q}}}{2}\big)
  \psi_i\big(\bm{q}+\frac{\tilde{\bm{q}}}{2}\big)}&=\int d\bm{p}\,e^{i\bm{p}\cdot\tilde{\bm{q}}/\hbar}\overline{W_i(\bm{p},\bm{q})}\\
  &={\mathcal{N}(E_i)}^{-1}(2\pi)^{d/2}mp_0^{d-2}\big(\frac{p_0\tilde{q}}{\hbar}\big)^{1-d/2}J_{d/2-1}\big(\frac{p_0\tilde{q}}{\hbar}\big),
  \label{SM:eq:two-point correlation}
  \end{aligned}
\end{equation}
where $p_0=\sqrt{2m[E_i-V(\bm{q})]}$.

Substituting Eq.~\eqref{SM:eq:two-point correlation} into Eq.~\eqref{SM:eq:offdiagonal variance init} and making the coordinate transformation
\begin{equation}
\begin{cases}
\bm{q}^{\,\prime\prime}
= \bm{q} - \dfrac{\tilde{\bm{q}}}{2}, \\[6pt]
\bm{q}^{\,\prime}
= \bm{q} + \dfrac{\tilde{\bm{q}}}{2}.
\end{cases}
\label{SM:eq:coornidate transformation}
\end{equation}
gives
\begin{equation}
  \begin{aligned}
  \overline{|\bra{E_i}O(\hat{\bm{q}})\ket{E_j}|^2}
  &=
  \hbar^{d-2}
  \bigg(\frac{2}{m}\bigg)^{d/2-1}
  \Gamma^2\bigg(\frac{d}{2}\bigg)
  \int d\bm{q}\,d\tilde{q}\,
  J_{d/2-1}\bigg(
  \frac{\sqrt{2m(E_i-V(\bm{q}))}\tilde{q}}{\hbar}
  \bigg)
  J_{d/2-1}\bigg(
  \frac{\sqrt{2m(E_j-V(\bm{q}))}\tilde{q}}{\hbar}
  \bigg)
  \\
  &\quad\times
  \tilde{q}
  \big[(E_i-V(\bm{q}))(E_j-V(\bm{q}))\big]^{
  \left(\frac{d}{4}-\frac{1}{2}\right)}
  \int d\bm{\theta}_{d-1}\,
  O(\bm{q}+\tilde{\bm{q}}/2)
  O(\bm{q}-\tilde{\bm{q}}/2)\\
  &\quad\times\bigg\{\int d\bm{q}\,[E_i-V(\bm{q})]^{d/2-1}\int d\bm{q}\,[E_j-V(\bm{q})]^{d/2-1}\bigg\}^{-1}.
  \end{aligned}
  \label{SM:eq:offdiagnal variance}
\end{equation}
Applying the asymptotic approximation for the Bessel functions, given in Eq.~\eqref{main:eq:J_cos} of the main text, yields
\begin{equation}
  \begin{aligned}
  \overline{|\bra{E_i}O(\hat{\bm{q}})\ket{E_j}|^2}
  &\approx
  \gamma{A(E_i)}^{-1}{A(E_j)}^{-1}\int d\bm{q}\,d\tilde{q}\,
  \\
  &\quad\times
  \bigg\{\cos\bigg(
  \frac{\sqrt{2m}\tilde{q}(E_i-E_j)}
  {\hbar\big(\sqrt{E_i-V(\bm{q})}
  +\sqrt{E_j-V(\bm{q})}\big)}
  \bigg)-\sin\bigg(\frac{\sqrt{2m}\tilde{q}}{\hbar}\big(\sqrt{E_i-V(\bm{q})}+\sqrt{E_j-V(\bm{q})}\big)-\frac{\pi d}{2}\bigg)\bigg\}
  \\
  &\quad\times
  \big[(E_i-V(\bm{q}))(E_j-V(\bm{q}))\big]^{
  \frac{d}{4}-\frac{3}{4}}
  \int d\bm{\theta}_{d-1}\,
  O(\bm{q}+\tilde{\bm{q}}/2)
  O(\bm{q}-\tilde{\bm{q}}/2).
  \end{aligned}
  \label{eq:offdiagonal variance approx}
\end{equation}
In the semiclassical limit $\xi_{sm}\to0$, the sine term oscillates rapidly
with the radial coordinate $\tilde{q}=|\tilde{\bm{q}}|$. Its contribution to
the integral is therefore suppressed by phase cancellation and can be
neglected at leading order. 

\begin{comment}
Figure~\ref{sm:fig:analytical_comparison} compares three expressions presented in the main text:
the full semiclassical result, Eq.~\eqref{main:eq:offdiagnal variance}, with
the large-argument Bessel approximation,
Eq.~\eqref{main:eq:offdiagonal variance approx}, and its low-frequency form,
Eq.~\eqref{main:eq:Oq_offdiag_Phi_approx_w}, after omitting corrections of
order $O\big((\omega/\overline{E})^2\big)$. The close agreement demonstrates
the accuracy of the successive approximations over the frequency range
shown.

\begin{figure}[htbp]
  \centering
  \includegraphics[width=0.72\textwidth]
  {figures/3analytical_compare.pdf}
  \caption{Comparison of three analytical predictions for the variance of
  the off-diagonal matrix elements of $q_x$ in the generalized quarter-Sinai billiard at
  $\overline{E}=10^4$. The solid orange curve is the full semiclassical
  result, Eq.~\eqref{main:eq:offdiagnal variance}; the dashed purple curve is
  the large-argument Bessel approximation,
  Eq.~\eqref{main:eq:offdiagonal variance approx}; and the blue star markers
  show the low-frequency approximation,
  Eq.~\eqref{main:eq:Oq_offdiag_Phi_approx_w}, with
  $O\big((\omega/\overline{E})^2\big)$ corrections omitted.}
  \label{sm:fig:analytical_comparison}
\end{figure}
\end{comment}

\section{Typical behavior of $\Phi(\tilde{\bm{z}})$}
Here we show the typical behavior of $\Phi(\tilde{\bm{z}})$, defined by
Eq.~\eqref{main:eq:Phi} in the main text, for the position operator $q_x$
in the generalized quarter-Sinai billiard.  The displacement is expressed in the
dimensionless coordinates
$\tilde{\bm{z}}=\tilde{\bm{q}}/\sqrt{S_\Omega}$.

\begin{figure}[htbp]
  \centering
  \includegraphics[width=0.4\textwidth]{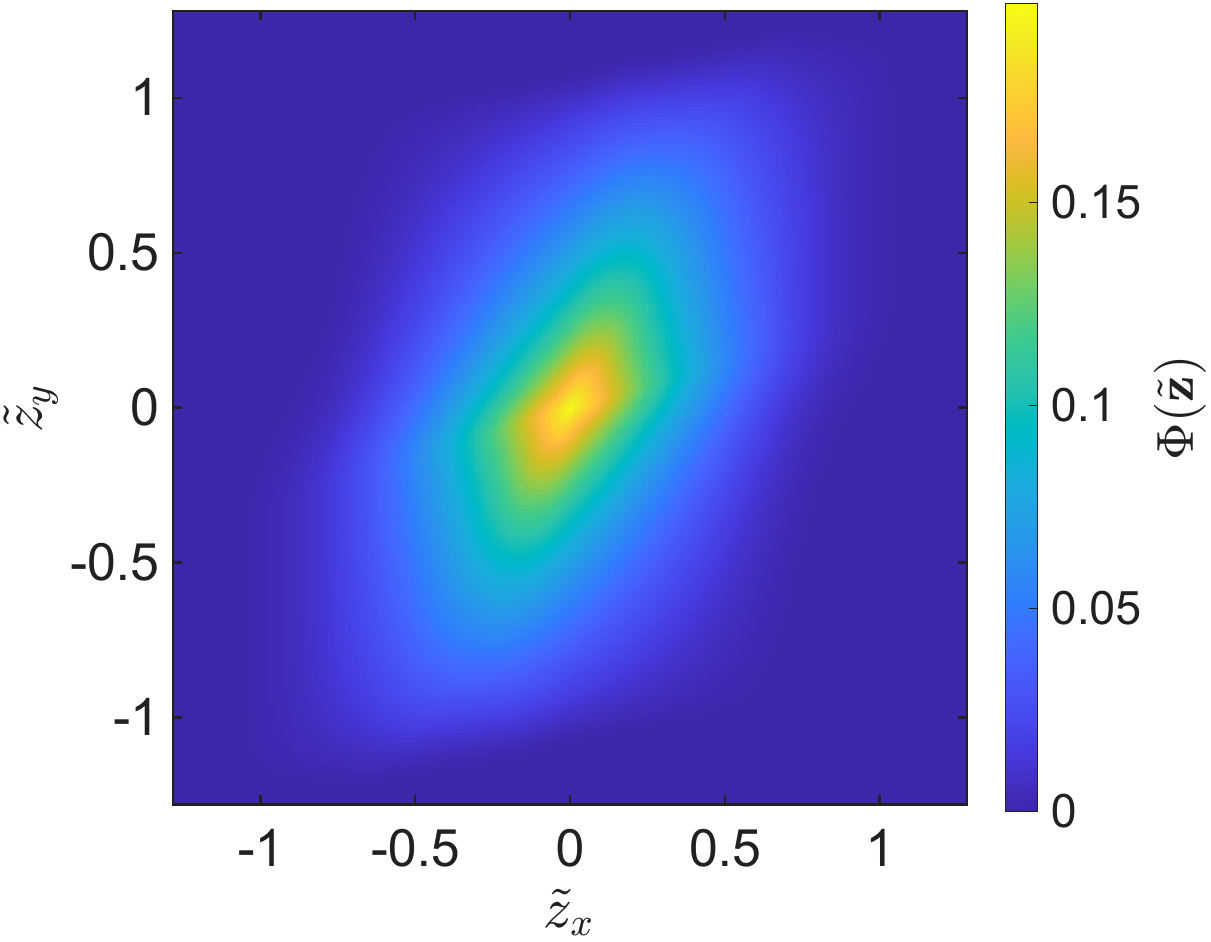}
  \caption{Numerically evaluated $\Phi(\tilde{\bm{z}})$ for the position
  operator $q_x$ in the generalized quarter-Sinai billiard.  The function is peaked at
  zero displacement and decays as the common integration domain shrinks.}
  \label{sm:fig:phi}
\end{figure}

For a nonzero displacement, the midpoint and both shifted points must all
remain inside the billiard.  The common integration domain therefore
contracts as $|\tilde{\bm{z}}|$ increases.  As shown in
Fig.~\ref{sm:fig:phi}, $\Phi(\tilde{\bm{z}})$ is maximal at the origin,
is invariant under $\tilde{\bm{z}}\to-\tilde{\bm{z}}$, and decreases toward
zero as the two shifted copies of the billiard cease to overlap.

\section{Derivation of the connected autocorrelation function}
\label{sm:sec:connected_autocorrelation}

We now derive the connected autocorrelation
function using the off-diagonal variance in
Eq.~\eqref{main:eq:Oq_offdiag_Phi_approx_w} in the main text.

The general relation between the connected autocorrelation function and $f_O(E,\omega)$ is \cite{D'Alessio03052016,PhysRevB.103.235137}
\begin{equation}
  C_O(t;E)
  =
  \int_{-\infty}^{\infty}d\omega\,
  e^{-i\omega t/\hbar}
  \left[
    \left|f_O(E,\omega)\right|^2
    +
    \frac{\omega}{2}
    \frac{\partial\left|f_O(E,\omega)\right|^2}{\partial E}
  \right].
  \label{sm:eq:correlation_gradient_expansion}
\end{equation}

For a two-dimensional billiard, the leading Weyl density of states is
\begin{equation}
  \rho(E)=\frac{m\SOmega}{2\pi\hbar^2},
  \label{sm:eq:billiard_density_of_states}
\end{equation}
which is independent of \(E\). We neglect higher-order corrections because we consider highly excited states. Thus the microcanonical inverse temperature, $\beta(E)=\partial_E\ln\rho(E)$, vanishes at this order, and
\begin{equation}
  \left|f_O(E,\omega)\right|^2
  =
  \rho(E)\,
  \overline{\left|
    \left\langle E_i\middle|O\middle|E_j\right\rangle
  \right|^2},
  \qquad
  E=\frac{E_i+E_j}{2}.
  \label{sm:eq:f_variance_relation}
\end{equation}
Using the low-frequency semiclassical variance derived in the main text,
we therefore find
\begin{equation}
  \left|f_O(E,\omega)\right|^2
  \simeq
  \mathcal A(E)
  \int d^2\widetilde{\bm z}\,
  \frac{\Phi(\widetilde{\bm z})}{\widetilde z}
  \cos\left[
    \kappa(E)\widetilde z\,\omega
  \right],
  \label{sm:eq:f_phi_relation}
\end{equation}
where
\begin{equation}
  \mathcal A(E)
  =
  \frac{\sqrt{mS_\Omega}}
  {2^{3/2}\pi^2\hbar\sqrt{E}},
  \qquad
  \kappa(E)
  =
  \frac{\sqrt{m\SOmega}}{\hbar\sqrt{2E}}.
  \label{sm:eq:A_gamma_definitions}
\end{equation}

For completeness, differentiating Eq.~\eqref{sm:eq:f_phi_relation} gives
\begin{equation}
\begin{split}
  \frac{\omega}{2}
  \frac{\partial\left|f_O(E,\omega)\right|^2}{\partial E}
  ={}&
  \mathcal A(E)\left\{
  -\frac{\omega}{4E}
  \int d\widetilde{\bm z}\,
  \frac{\Phi(\widetilde{\bm z})}{\widetilde z}
  \cos\left[\kappa(E)\widetilde z\,\omega\right]
  +\frac{\kappa(E)\omega^2}{4E}
  \int d\widetilde{\bm z}\,
  \Phi(\widetilde{\bm z})
  \sin\left[\kappa(E)\widetilde z\,\omega\right]
  \right\}.
\end{split}
\label{sm:eq:f_energy_derivative}
\end{equation}
Both terms on the right-hand side are odd functions of \(\omega\).
Consequently, they contribute only to the imaginary part of \(C_O(t;E)\) that is
antisymmetric under \(t\to-t\).

We define the symmetrized connected autocorrelation function as
\begin{equation}
  C_O^{\mathrm{sym}}(t;E)
  =
  \frac{C_O(t;E)+C_O(-t;E)}{2}=
  \frac{1}{2}
  \left\langle
    \left\{\delta O(t),\delta O(0)\right\}
  \right\rangle_E .
\label{sm:eq:symmetric_correlation_definition}
\end{equation}
Here $\delta O(t)=O(t)-\langle O\rangle_E$ denotes the fluctuation of the observable about its microcanonical mean.
Equation~\eqref{sm:eq:correlation_gradient_expansion} then reduces to
\begin{equation}
  \begin{split}
  C_O^{\mathrm{sym}}(t;E)
  &={}
  \mathcal A(E)
  \int d\widetilde{\bm z}\,
  \frac{\Phi(\widetilde{\bm z})}{\widetilde z}
  \int_{-\infty}^{\infty}d\omega\,
  \cos\left(\frac{\omega t}{\hbar}\right)
  \cos\left[\kappa(E)\widetilde z\,\omega\right]\\
  &  =
  \frac{1}{2\pi}
  \int d\tilde{z}d\theta\,
  \Phi\left(\tilde{\bm{z}}\right)\delta\left(\tilde{z}-\sqrt{\frac{2E}{mS_\Omega}}|t|\right).
\end{split}
\label{sm:eq:symmetric_correlation_transform}
\end{equation}

Writing \(\widetilde{\bm z}=(\widetilde z,\theta)\) in polar
coordinates and carrying out the radial integral yields
\begin{equation}
  C_O^{\mathrm{sym}}(t;E)
  =
  \frac{1}{2\pi}
  \int_0^{2\pi}d\theta\,
  \Phi\left(
    \frac{v(E)\lvert t\rvert}{\sqrt{\SOmega}},
    \theta
  \right)
  ,
  \label{sm:eq:symmetric_correlation_final}
\end{equation}
where \(v(E)=\sqrt{2E/m}\) is the classical particle speed. Thus, the
symmetrized connected autocorrelation function is proportional to the angular
average of the spatial overlap function evaluated at the dimensionless
displacement \(v(E)|t|/\sqrt{S_\Omega}\). Because $\Phi(\tilde{\bm z})$ is
localized near the origin, the correlation function decays on the characteristic
classical timescale $t\sim\sqrt{S_\Omega}/v(E)$.

\bibliography{references}